# Communication via FRET in Nanonetworks of Mobile Proteins

Jakub Kmiecik
Department of Telecommunications,
AGH University of Science and Technology
Al. Mickiewicza 30
30-059 Krakow, Poland
jkmiecik@agh.edu.pl

Pawel Kulakowski
Department of Telecommunications,
AGH University of Science and Technology
Al. Mickiewicza 30
30-059 Krakow, Poland
kulakowski@kt.agh.edu.pl

Krzysztof Wojcik
Division of Cell Biophysics,
Faculty of Biochemistry,
Biophysics and Biotechnology,
Jagiellonian University
7 Gronostajowa St.
30-387 Krakow, Poland
krzysztof.wojcik@uj.edu.pl

Andrzej Jajszczyk
Department of Telecommunications,
AGH University of Science and Technology
Al. Mickiewicza 30
30-059 Krakow, Poland
jajszczyk@kt.agh.edu.pl

## ABSTRACT
A practical, biologically motivated case of protein complexes (immunoglobulin G and FcRII receptors) moving on the surface of leukocyte cells, that are common parts of an immunological system, is investigated. These proteins are considered as nanomachines creating a large nanonetwork. Accurate molecular models of the proteins and the fluorophores which act as their nanoantennas are used to simulate the communication between the nanomachines when they are close to each other. The theory of diffusion-based Brownian motion is applied to model movements of the proteins. It is assumed that fluorophore molecules send and receive signals using the Förster Resonance Energy Transfer. The probability of the efficient signal transfer, the respective bit error rate, and the communication channel capacity are calculated and discussed.

## 1. INTRODUCTION
Nanocommunication is a rapidly expanding area of communication sciences, trying to facilitate the development of nanotechnology. Future nanorobots and nanomachines will have to communicate with each other, so there is an urgent need for designing communication schemes proper for such tiny devices. In the recent years, some techniques have been proposed, most of them based on the following two approaches: scaling down the technical solutions existing in wireless and wired communications or adapting the ones already present in biological systems. Most of these techniques are, however, still not feasible to be realized in nanoscale (miniaturized transceivers and antennas) or they are very slow because of their considerable propagation delay (calcium signaling, moving bacteria, molecular motors).

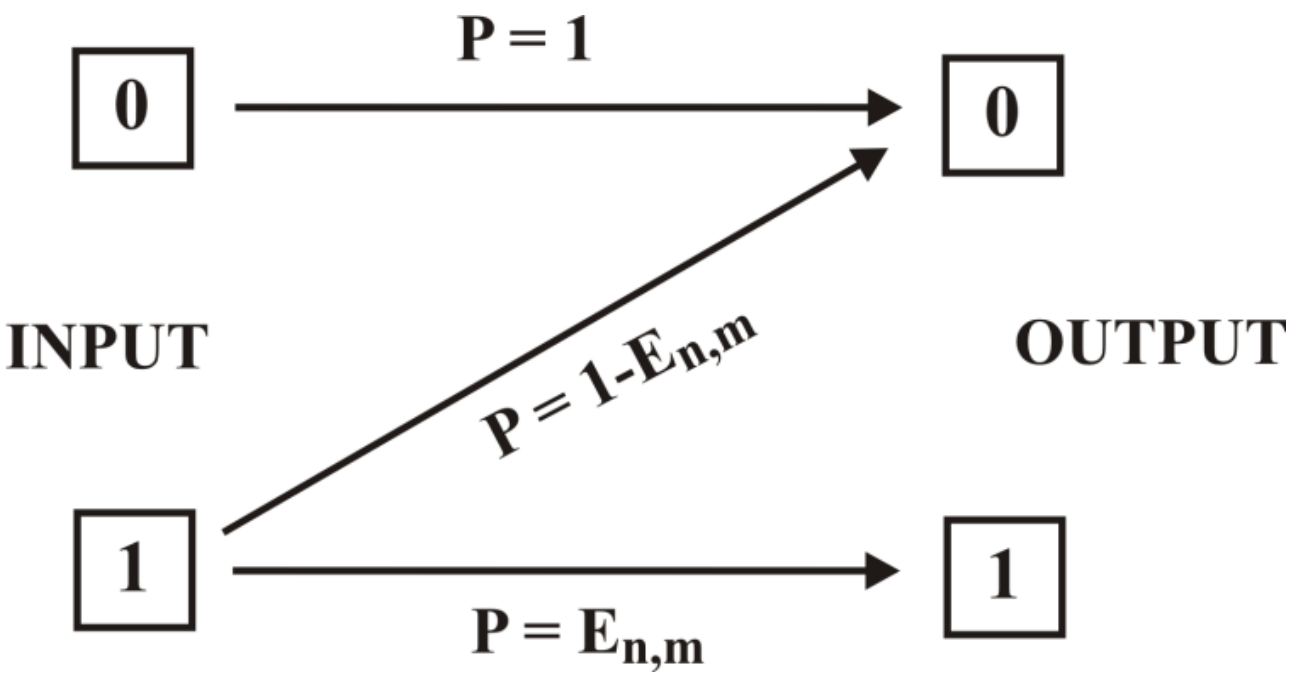

**Fig. 1. FRET communication channel and transmission probabilities for bits '0' and '1'.**

One of the most promising nanoscale communication techniques is based on Förster Resonance Energy Transfer (FRET). The FRET phenomenon for the nanocommunication purposes was proposed a few years ago [11, 7, 6] and studied both theoretically and experimentally [15, 8, 13]. FRET is characterized by small propagation delays, usually few dozens of nanoseconds, and allows for transmissions of data streams with bit rates of several Mbit/s. It has also been recently shown that the main drawback of FRET, which is its quite high bit error rate (BER), may be decreased by using multiple nanoantennas, creating the so called MIMO-FRET communication channels [15].

In this paper, we investigate a scenario of mobile nanonodes communicating via FRET. While a theoretical analysis of mobile FRET-based nanonetworks has been already presented in [5], here we investigate a practical, biologically motivated case of protein complexes (immunoglobulin G molecules and their receptors) moving on the surface of a leukocyte cell. Leukocytes are common parts of an immunological system of each human or animal body. There are, on average, 40 thousand protein complexes on the surface of each cell. These proteins constantly move fueled by Brownian motion; they approach each other, creating opportunities for communication. We consider them as nanomachines creating a large nanonetwork. We use accurate molecular models of the proteins and the fluorophores which act as their nanoantennas. With these models, we simulate the communication between the nanomachines when they are moving close to each other. We calculate the probability of the efficient signal transfer, the respective bit error rate, and the communication channel capacity.

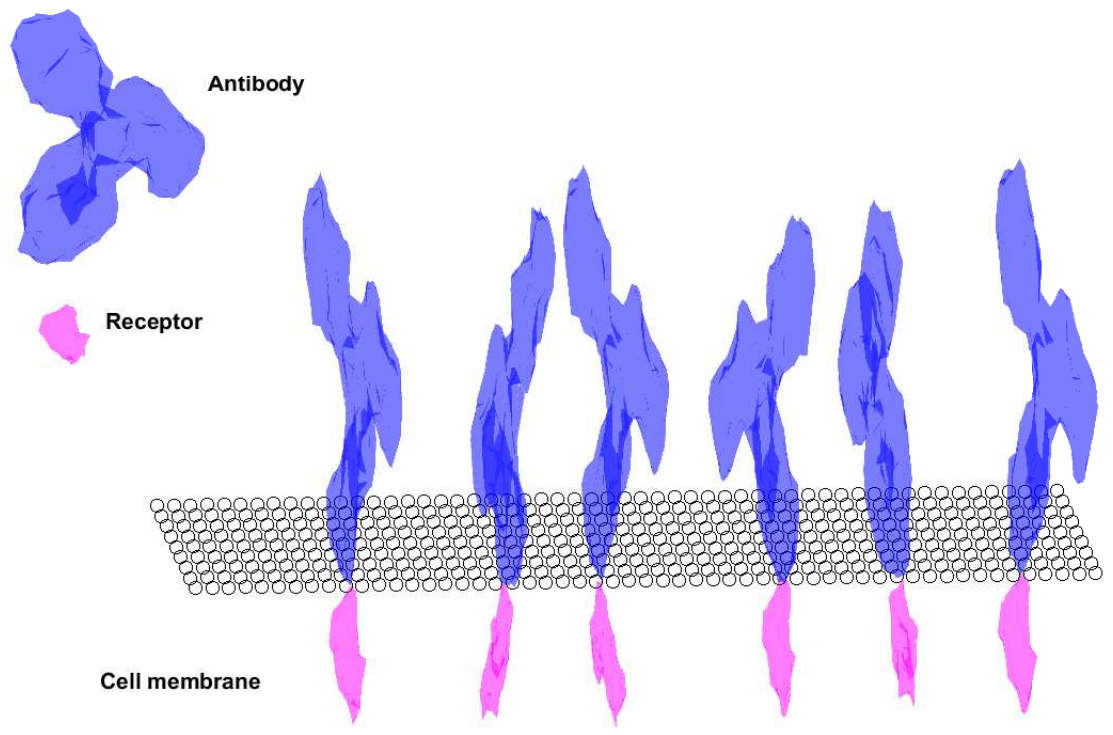

**Fig. 2. Antibodies with attached Fc receptors.**

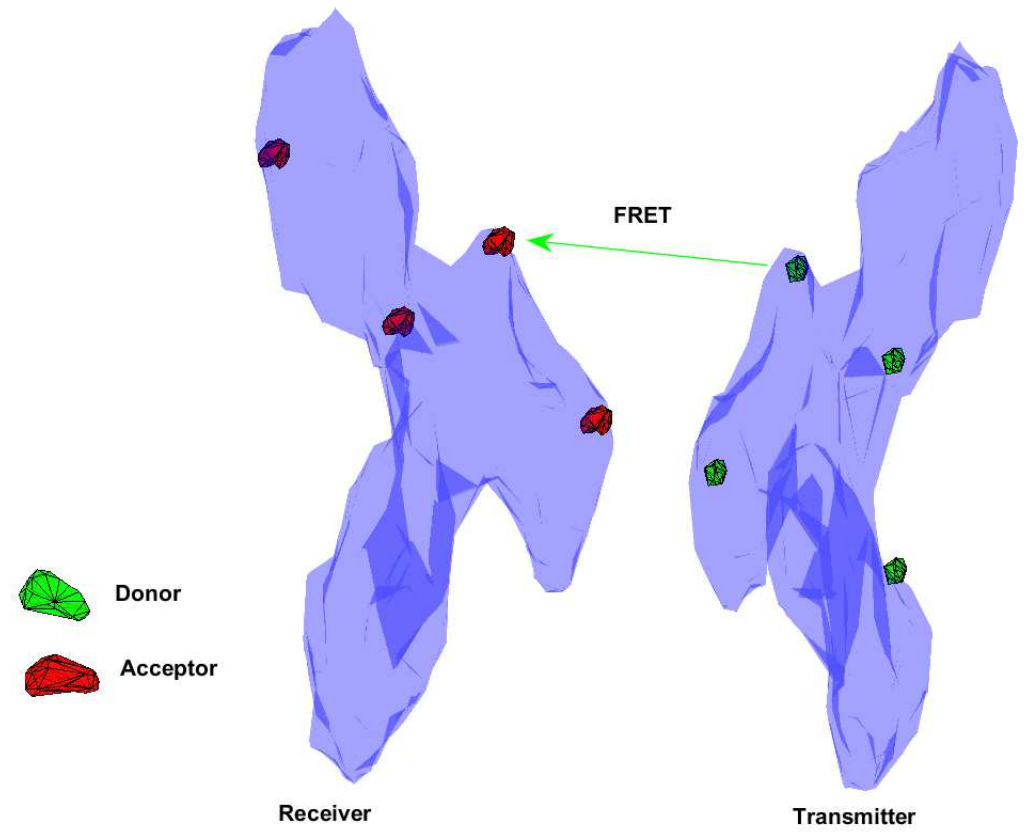

**Fig. 3. Fluorophores mounted on antibodies.**

The rest of the paper is organized as follows. In Section 2, the basics of the FRET-based communication are introduced together with the FRET channel model. The biological aspects of the analyzed mobile nanonetworks are presented in Section 3. The details of the simulation scenario are provided in Section 4. In Section 5, the results of the FRET efficiency are shown. In Section 6, the respective bit error rates and channel capacity are calculated. Finally, Section 7 concludes the paper.

## 2. FRET BASICS

The Förster Resonance Energy Transfer is a physical phenomenon where an excited molecule, called donor, non-radiatively passes its energy to another molecule, called acceptor, located in its vicinity. Both molecules must be spectrally matched, i.e., the donor emission spectrum should overlap, at least partially, with the acceptor absorption spectrum. Such a pair of molecules may be seen as a wireless communication system: the donor is a transmitter antenna and the acceptor plays the role of the receiver antenna. In general, if the donor is excited, it may return to its ground state in two ways: emitting a photon or via FRET. The probability of FRET is called FRET efficiency and is given by the following formula:

$$E = \frac{R_0^6}{r^6 + R_0^6} \tag{1}$$

where $r$ is the donor-acceptor separation and $R_0$ is the so-called Förster distance which is a parameter characteristic for each pair of two molecules, depending on their emission/absorption spectra. The FRET efficiency strongly depends on the separation between the donor and the acceptor: if $r = R_0$, $E_{FRET}$ equals 50% and it decreases with the sixth power of that distance.

The probability of successful transmission may be increased if there are more molecules involved, i.e., more donors and acceptors at both sides of the communication channel. Having all the donors excited at the same time, it is enough that one of them passes its energy to an acceptor, then the signal is successfully sent via the communication channel. The FRET efficiency with one donor and $m$ acceptors is known as [15]:

$$E_{1,m} = \frac{R_0^6 \sum_{i=1}^{m} \frac{1}{r_i^6}}{1 + R_0^6 \sum_{i=1}^{m} \frac{1}{r_i^6}} \tag{2}$$

where $r_i$ is the separation between the donor and the $i$-th acceptor. Now, having $n$ donors excited at once, we would like to calculate the probability that *at least one* donor passes its energy to any acceptor via FRET. We can use the probability of the complementary event, i.e., none of the donors passes its energy via FRET, which is:

$$\left(1 - \frac{R_0^6 \sum_{i=1}^{m} \frac{1}{r_{1i}^6}}{1 + R_0^6 \sum_{i=1}^{m} \frac{1}{r_{1i}^6}}\right) \cdots \left(1 - \frac{R_0^6 \sum_{i=1}^{m} \frac{1}{r_{ni}^6}}{1 + R_0^6 \sum_{i=1}^{m} \frac{1}{r_{ni}^6}}\right) \tag{3}$$

where $r_{ki}$ is the separation between the $k$-th donor and the $i$-th acceptor. Finally, the probability of at least one FRET process in the ($n$, $m$) FRET channel is given by:

$$E_{n,m} = 1 - \prod_{k=1}^{n} \left(1 - \frac{R_0^6 \sum_{i=1}^{m} \frac{1}{r_{ki}^6}}{1 + R_0^6 \sum_{i=1}^{m} \frac{1}{r_{ki}^6}}\right) \tag{4}$$

This probability can be understood as the FRET ($n$, $m$) efficiency.

The FRET process may be used to create a communication channel. When bit ‘0’ is going to be sent, it is realized by keeping the donors in the ground state, so there is no transmission. Thus, ‘0’ is always sent successfully. On the other hand, when sending bit ‘1’, all the donors are excited and the transmission is successful with the probability $E_{n,m}$. This channel and the probabilities of correct and erroneous transmissions of both bits are illustrated in Fig. 1. The delay of the FRET process is very small, 10-20 nanoseconds or even less. When sending a bit sequence, one must, however, wait after each transmission to have the acceptors (receivers) release their energy. It is thus safe to assume that a bit may be transmitted once per 40 nanoseconds, which results in the maximal throughput of 25 Mbit/s [13].

## 3. THE MOBILE FRET NETWORK

As explained in the previous section, the FRET efficiency highly depends on the distance: it decreases with the sixth power of the donor-acceptor separation. Thus, nano-machines that would like to communicate via FRET should approach very close each other, at least for a time period suitable for data transmission. Such a situation may happen in a mobile nanonetwork where the nanonodes have capabilities to move through their environment.

In this paper, we consider a scenario of a mobile nanonetwork motivated by a real biological situation. We analyze antibodies moving on the surface of a mast cell. Mast cells are common parts of an immunological system of each human or animal body. These cells, having the average diameter of about 10 micrometers, are covered by a lipid membrane. There are, on average, 40 thousand of antibodies (immunoglobulin) on the surface of each mast cell. Each antibody is attached to an Fc receptor, which can, however, float freely in the lipid membrane (Fig. 2).

As we described in our previous works [15, 13], antibodies are very well suited to act as nanomachines. They may perform several functions in living organisms, recognizing and binding other molecules. Smaller molecules called fluorophores may be mounted on the antibodies (in life sciences, the process of attaching fluorophores to antibodies is called *labeling*). The fluorophores serve as nanoantennas and communicate with other fluorophores mounted on other antibodies via FRET (Fig. 3).

The antibodies together with Fc receptors move over the surface of the lipid membrane fueled by Brownian motion. As Brownian motion is a random process, it cannot be exactly predicted if two specific antibodies approach each other during their movement. Having in mind a quite high density of antibodies on the lipid membrane, about 30 molecules per square micrometer, we can, however, assume that antibodies are frequently coming very close to each other. These moments of closeness enable for an exchange of information than can be further routed through the whole nanonetwork using, e.g., opportunistic communication schemes [12].

In this paper, we focus our study on a pair of antibodies located close enough to each other that the FRET communication is possible. Because of the randomness of the antibodies movement, we cannot predict the exact moment of time when they approach each other. Instead, we track their movement when they are already in a close distance and they move freely, via Brownian motion, which finally makes them separated. We calculate the efficiency of the FRET transmission, the channel bit error rate, and, finally, the capacity of the communication channel.

## 4. SIMULATION SCENARIO

The analysis of the scenario described above has been performed via simulations by using suitable structural models of all the network components. The molecular structures of antibodies, immunoglobulin G (acting as our nanomachines), have been taken from Protein Data Bank, using the 1IGT model [14]. These antibodies are about 15 nm long, with a characteristic shape of 3-element airscrew (Fig. 3). As fluorophores (nanoantennas), we have chosen Atto 610 dyes as donors and Atto 655 dyes as acceptors. The Förster distance between Atto 610 and Atto 655 is quite large, i.e., 7.6 nm, which is very advantageous for efficient FRET transmission. Fluorophores are much smaller particles, with the diameter of about 1.5 nm. The models of both fluorophores are available at PubChem [2, 3]. Both Protein Data Bank and PubChem are chemical databases providing free access to their content.

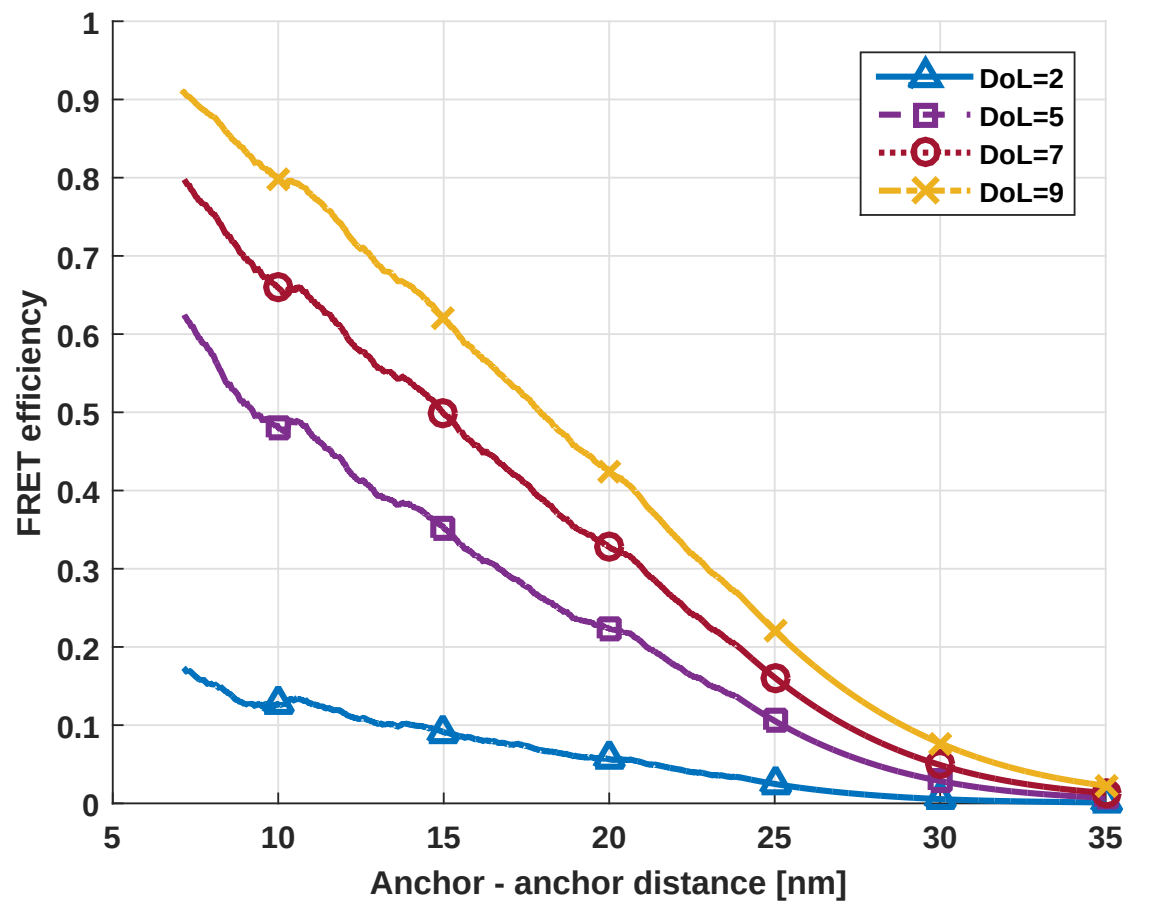

**Fig. 4. FRET efficiency for two communicating nanomachines as a function of their separation. Degree of labeling (DoL) of both nanomachines ranges from 2 to 9.**

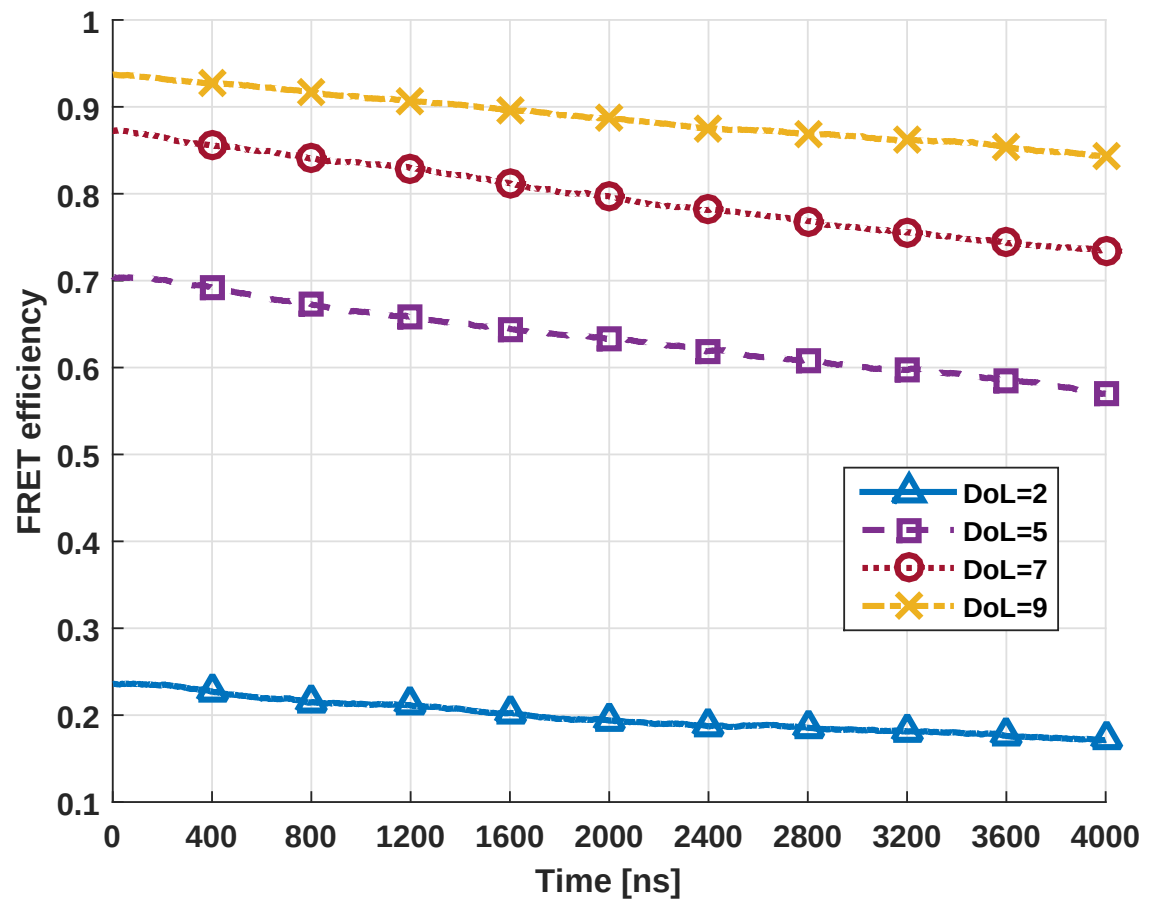

**Fig. 5. FRET efficiency for two communicating nanomachines moving via Brownian motion. Degree of labeling (DoL) of both nanomachines ranges from 2 to 9.**

The typical degree of labeling (the number of donors/acceptors per antibody) for Atto dyes ranges from 2 to 9 [4] and these values were simulated. The bonds between the fluorophores and antibodies are of NHS ester type [1], therefore, analyzing the molecular structure of immunoglobulin G we could state there were 25 possible binding positions of fluorophores on the antibody. The exact positions were chosen randomly in each simulation run.

Each immunoglobulin G antibody was attached to its suitable receptor, which was FcRII [9]. Receptors were working as carriers moving the antibodies around; they were floating in the lipid membrane fueled by Brownian motion. The movements of each complex of the FcRII receptor and antibody labeled with Atto fluorophores was simulated according to the formula describing Brownian motion [10]:

$$B(t_2) - B(t_1) \sim N\left(0, \sigma^2 (t_2 - t_1)\right) \quad (5)$$

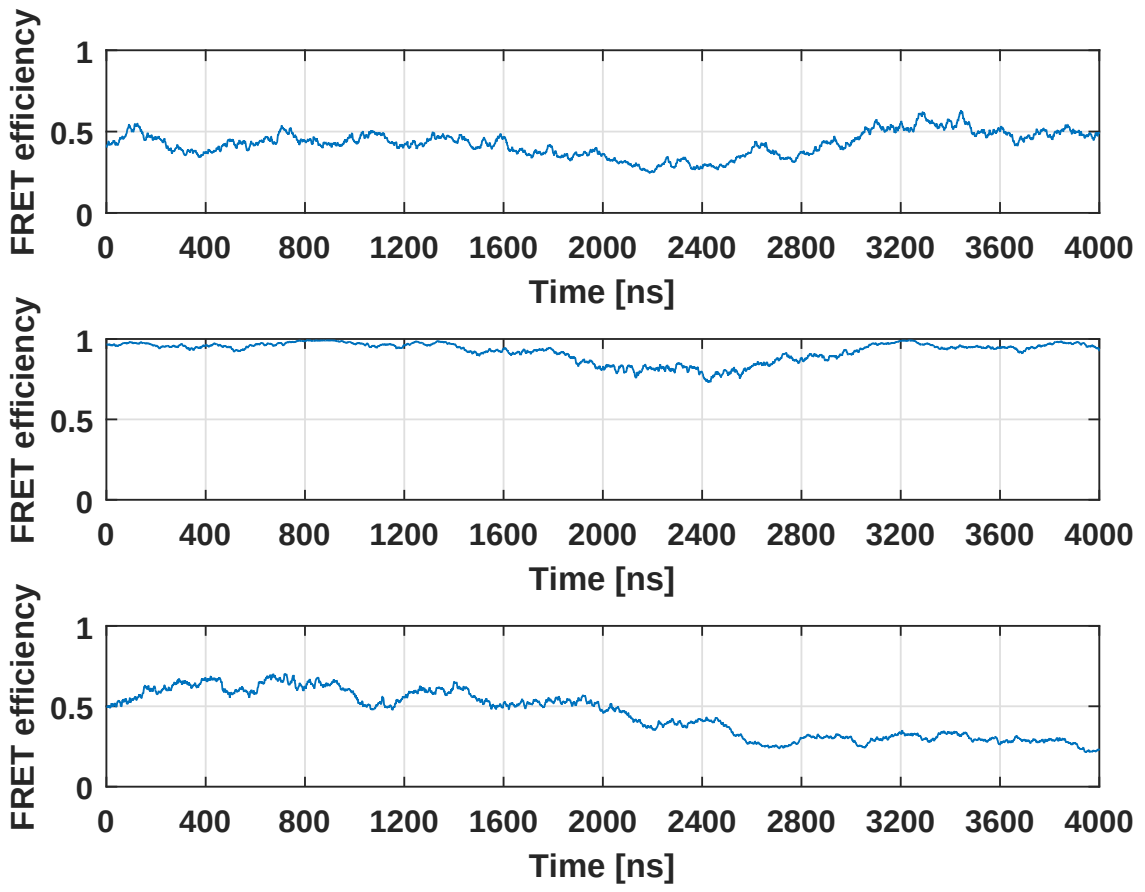

**Fig. 6. Examples of FRET efficiency changes as a function of time.**

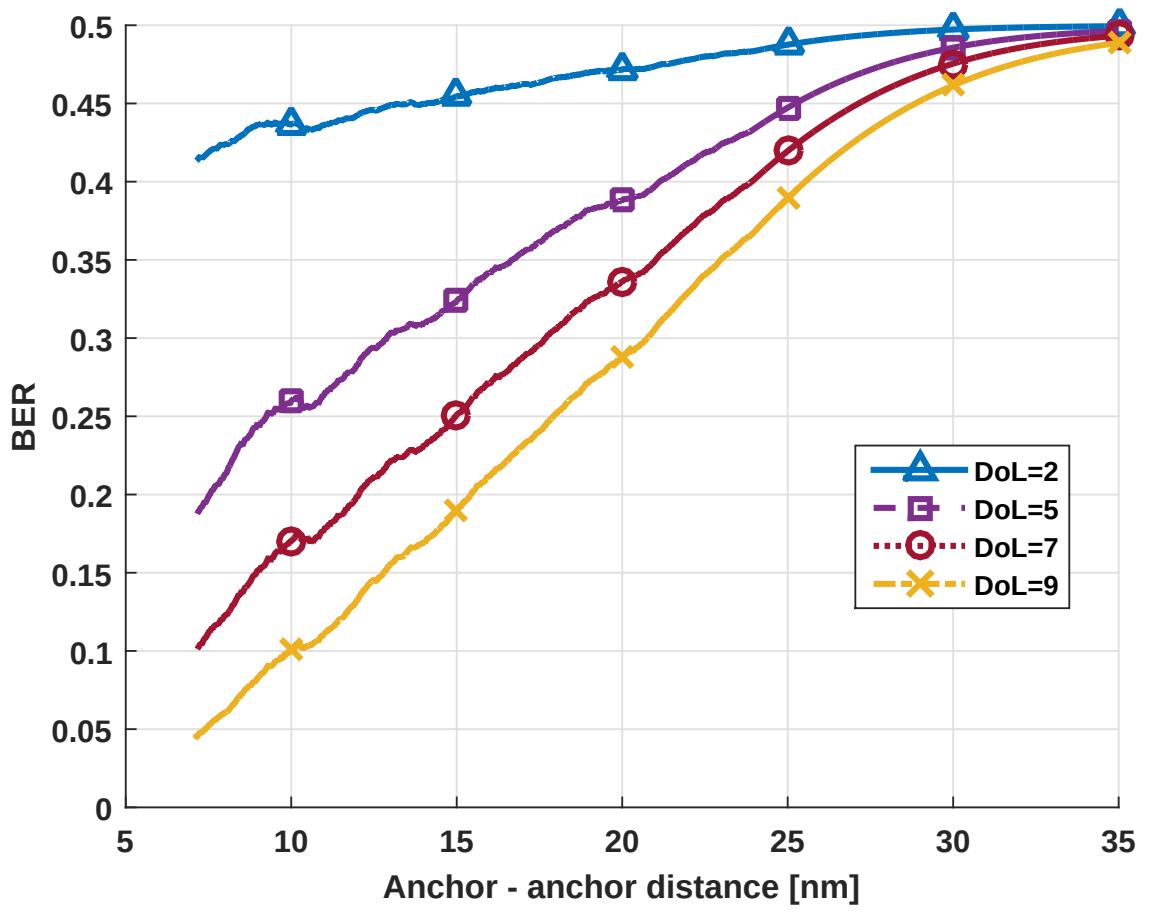

**Fig. 7. Bit error rate for two communicating nanomachines as a function of their separation. Degree of labeling (DoL) of both nanomachines ranges from 2 to 9.**

In the formula above, $B(t_1)$, $B(t_1)$ represent positions of an antibody at times $t_1$, $t_2$ respectively, so $B(t_1)$- $B(t_2)$ is a shift of the antibody in the time interval $t_2$-$t_1$. The shift is a Gaussian random variable with 0 mean and $\sigma^2(t_2\text{-}t_1)$ variance. In Brownian motion, $\sigma^2$ is given by:

$$\sigma^2 = \alpha D \tag{6}$$

where $D$ is the free diffusion coefficient of the considered particle propagating in the specific medium and $\alpha$ is 1, 2 or 3 in 1-, 2- or 3-dimentional system, respectively.

# 5. FRET EFFICIENCY RESULTS

The purpose of the conducted FRET efficiency simulations was twofold. First, we measured how the FRET efficiency depended on the average distance between the communicating nanomachines. The nanomachines considered here, i.e., immunoglobulin G antibodies with Atto fluorophores, were irregular molecular structures (see Fig. 3), so their relative orientation on the lipid membrane and the exact positions of the fluorophores on antibodies had a significant impact on the FRET efficiency. The simulations were thus conducted 1000 times; each time the nanomachines orientation and fluorophores positions were chosen randomly. The distance between the nanomachines was calculated as a separation between the centers of the Fc receptors where the antibodies were attached. For close distances, smaller than 15.8 nm, there were cases where antibodies overlapped each other. Such cases were obviously removed from the calculations. The results, i.e., the average FRET efficiency as a function of the distance between the nanomachines, are presented in Fig. 4.

Second, we investigated *how long* the communication might take place. Assuming that two antibodies were already very close to each other (just 1 nm of surface-to-surface separation), we simulated their movements with Brownian motion and calculated how the FRET efficiency depended on time. The simulation results for different degrees of labeling are given in Fig. 5; they are averaged over 1000 simulation runs. For comparison, in Fig. 6, three examples of single runs are presented. We can see that large deviations in FRET efficiency may occur even during 200 nanoseconds which corresponds with the time of sending only 5 bits (see Section 2).

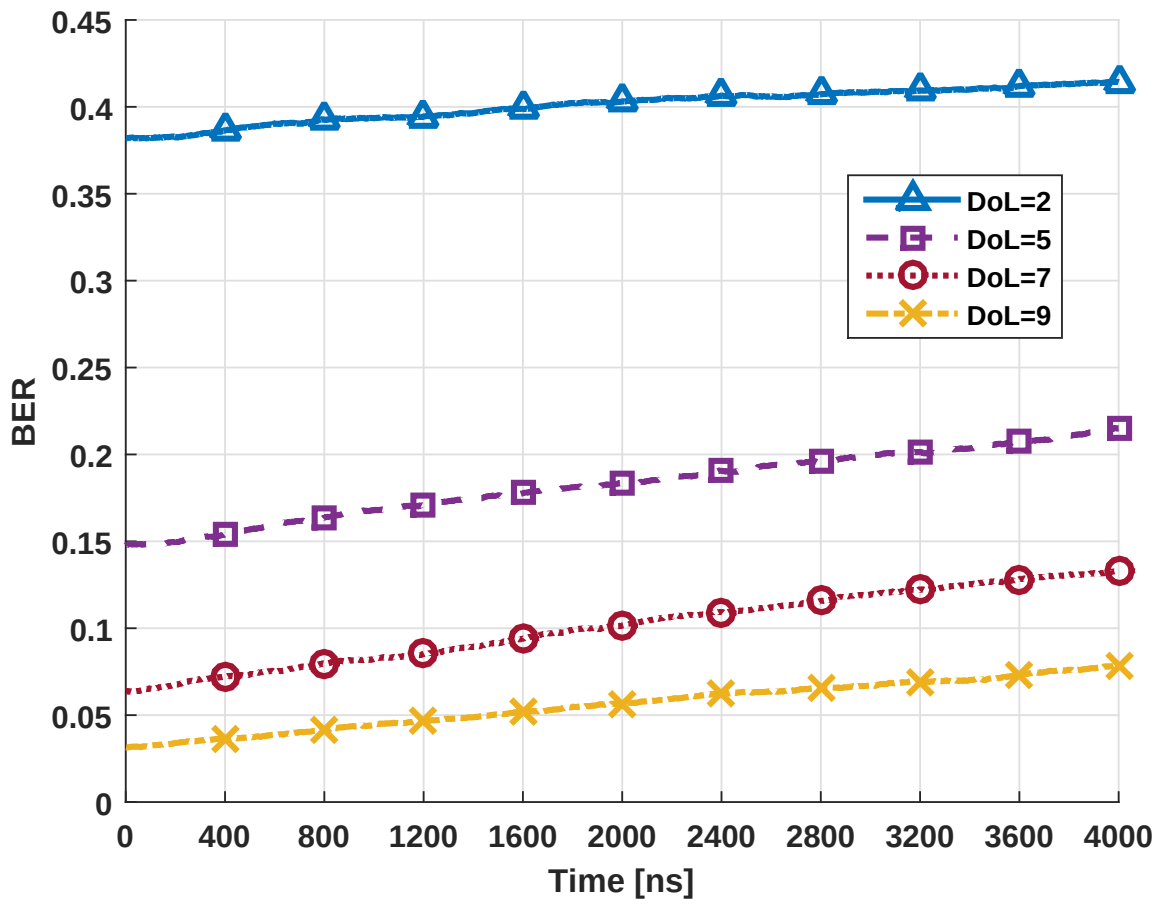

**Fig. 8. Bit error rate for two communicating nanomachines moving via Brownian motion. Degree of labeling (DoL) of both nanomachines ranges from 2 to 9.**

# 6. BER AND CHANNEL CAPACITY

While FRET efficiency is a physical parameter informing us how successful the FRET phenomenon is at transmitting signals, we would like to measure more parameters of the communication channel, i.e., the bit error rate and the channel capacity.

The channel bit error rate may be easily calculated on the basis of the channel model given in Section 2 and the given FRET efficiency. The channel is not symmetric (see Fig. 1): bit '0' is always transmitted correctly, while transmitting bit '1' is erroneous with the probability of $1 - E$. Thus, assuming that transmissions of '0' and '1' are equally probable, the bit error rate is equal to:

$$\text{BER} = 0.5(1 - E) \tag{7}$$

Therefore, having the statistics of FRET efficiency (Figs. 4 and 5), we may also present the respective bit error rate curves (Figs. 7 and 8). In Fig. 7, the bit error rate as a function of the distance between the nanomachines is given, while in Fig. 8 we show the BER statistics for the case of two nanomachines moving with Brownian motion. From these figures, we can see that BER values typical for telecommunication systems, i.e., below $10^{-3}$, are really hard to obtain in real nanoscale scenarios. In the best case, when the degree of labeling is high, BER is about $10^{-1}$. It has an impact on coding schemes that could be applied: the codes should probably contain more redundancy (a detailed analysis of suitable coding techniques is however outside of scope of this paper).

Finally, we analyzed the relative channel capacity. Relative capacity is calculated with reference to its highest possible value, which was assessed as 25 Mbit/s (see Section 2). It can be calculated as the mutual information of channel input and output, maximized over all possible input distributions:

$$C = \max_{p(x)} \sum_{y \in Y} \sum_{x \in X} p(x,y) \log_2 \left( \frac{p(x,y)}{p(x)p(y)} \right) \qquad (8)$$

where $X$ and $Y$ are channel input and output random variables, $p(x)$ and $p(y)$ are their distributions, and $p(x,y)$ is their joint distribution. For the FRET communication channel (see Fig. 1), we can easily calculate the probability distributions $p(x)$, $p(y)$, and $p(x,y)$ as functions of $E$ and $P_0$ (which is the probability that the input is '0') and then the channel capacity.

The relative channel capacity is illustrated in Fig. 9. The capacity clearly increases with the FRET efficiency (the red dashed curve) and it approaches 1 with $E \to 1$. The blue solid curve shows how the optimal $P_0$ depends on $E$: for $E \to 1$; the optimal $P_0$ approaches 0.5, i.e., inputs '0' and '1' should be equally probable. For low values of $E$, the optimal $P_0$ is however higher, reaching about 0.6 - 0.65, which means that input '0's is more favorable: it is intuitive, as '0's are transmitted without errors.

While controlling the value of $P_0$ is possible with a proper source encoder, in practice it is almost unfeasible, as the FRET efficiency may change very frequently (see Fig. 6 for some examples). It would be much more practical to have $P_0$ fixed. The green dotted curve in Fig. 9 shows that keeping $P_0 = 0.5$ is a quite good choice: the relative channel capacity is very close to the maximal one.

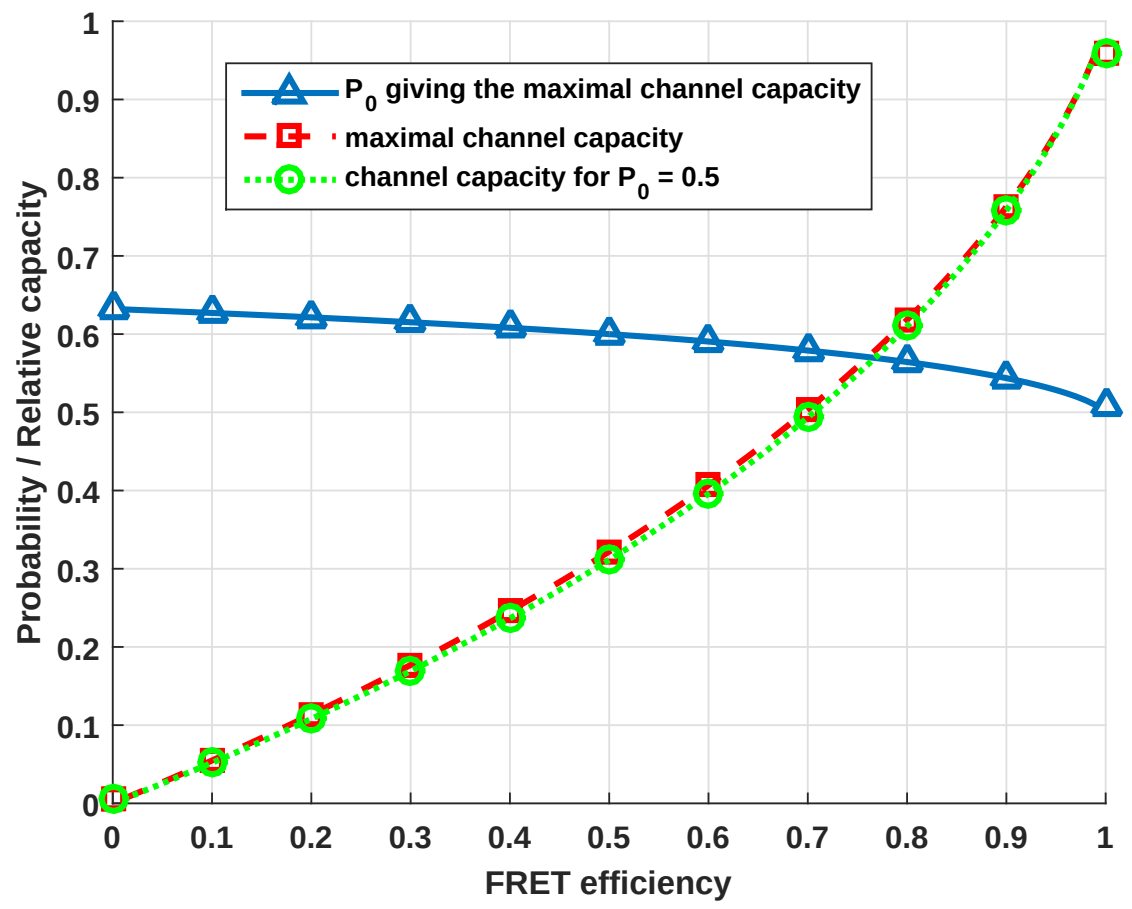

**Fig. 9. The channel capacity as a function of FRET efficiency in two cases: for $P_0 = 0.5$ (green dotted curve) and for the optimal $P_0$ (red dashed curve). The value of the optimal $P_0$ is additionally presented as a blue solid curve.**

## 7. CONCLUSIONS

In this paper, we analyzed a biologically motivated case of mobile nanonetworks. We considered a network consisting of protein complexes (immunoglobulin G, each about 15 nm long and its receptor FcRII) moving via Brownian motion on the surface of a leukocyte cell. The moving proteins occasionally closely approach each other, creating opportunities for communication. There are fluorophores (smaller molecules, each about 1.5 nm of diameter) attached to each protein, acting as nanoantennas. Fluorophores may send and receive signals exploiting Förster Resonance Energy Transfer as a physical phenomenon.

Using molecular models of proteins and fluorophores, as well as the theory of diffusion-based Brownian motion, we simulated the movements of proteins and the communication between them. We calculated the FRET efficiency which expresses the information transfer efficacy and then the respective bit error rate and channel capacity. The results show that the transmission with the throughput of several Mbit/s may be maintained over a few thousand of nanoseconds, especially when the transmitting and receiving nanomachines (proteins) are equipped with large number of nanoantennas (fluorophores).

## 8. REFERENCES

[1] Amine-reactive Crosslinker Chemistry, *ThermoFisher technical paper*, https://www.thermofisher.com/pl/en/home/life-science/protein-biology/protein-biology-learning-center/protein-biology-resource-library/pierce-protein-methods/amine-reactive-crosslinker-chemistry.html, accessed 31 March 2016.

[2] Atto 610 N-succinimidyl ester model: https://pubchem.ncbi.nlm.nih.gov/compound/16218776, *PubChem Open Chemistry Database*, accessed 31 March 2016.

[3] Atto 655 NHS ester model: https://pubchem.ncbi.nlm.nih.gov/compound/16218518, *PubChem Open Chemistry Database*, accessed 31 March 2016.

[4] Atto 655 Protein Labeling Kit – Product Information: https://www.sigmaaldrich.com/content/dam/sigma-aldrich/docs/Sigma/Datasheet/6/73919dat.pdf, *Sigma-Aldrich Co. LLC*, the web page accessed 31 March 2016.

[5] M. Kuscu, O. B. Akan, A Communication Theoretical Analysis of FRET-Based Mobile Ad Hoc Molecular Nanonetworks, *IEEE Transactions on NanoBioscience*, vol. 13, no. 3, pp. 255-266, September 2014.

[6] M. Kuscu, O. B. Akan, Multi-Step FRET-Based Long-Range Nanoscale Communication Channel, *IEEE Journal on Selected Areas in Communications*, vol. 31, no. 12, pp. 715-725, December 2013.

[7] M. Kuscu, O. B. Akan, A Physical Channel Model and Analysis for Nanoscale Communications with Förster Resonance Energy Transfer (FRET), *IEEE Transactions on Nanotechnology*, vol. 11, no. 1, pp. 200-207, January 2012.

[8] M. Kuscu, A. Kiraz, O. B. Akan, Fluorescent Molecules as Transceiver Nanoantennas: The First Practical and High-Rate Information Transfer over a Nanoscale Communication Channel based on FRET, *Nature Scientific Reports*, vol. 5, no. 7831, January 2015.

[9] J. Michl, M. M. Pieczonka, J. C. Unkeless, G. I. Bell, S. C. Silverstein, Fc receptor modulation in mononuclear phagocytes maintained on immobilized immune complexes occurs by diffusion of the receptor molecule, *The Journal of Experimental Medicine*, vol. 157, no. 6, pp. 2121-2139, 1983.

[10] T. Nakano, A. W. Eckford, T. Haraguchi, *Molecular Communication*, Cambridge University Press, New York, 2013.

[11] L. Parcerisa, I. F. Akyildiz, Molecular Communication Options for Long Range Nanonetworks, *Computer Networks (Elsevier) Journal*, vol. 53, pp. 2753-2766, 2009.

[12] L. Pelusi, A. Pasarella, M. Conti, Opportunistic Networking: Data Forwarding in Disconnected Mobile Ad Hoc Networks, *IEEE Commnications Magazine*, vol. 44, no. 11, pp. 134-141, November 2006.

[13] K. Solarczyk, K. Wojcik, P. Kulakowski, Nanocommunication via FRET with DyLight Dyes using Multiple Donors and Acceptors, *IEEE Transactions on NanoBioscience*, accepted for publication.

[14] Structure of immunoglobulin G – 1IGT model: http://www.rcsb.org/pdb/explore/explore.do?structureId=1IGT, *RCSB Protein Data Bank*, accessed 31 March 2016.

[15] K. Wojcik, K. Solarczyk, P. Kulakowski, Measurements on MIMO-FRET nanonetworks based on Alexa Fluor dyes, *IEEE Transactions on Nanotechnology*, vol. 14, no. 3, pp. 531-539, May 2015.